\documentclass[]{spie}  

\usepackage{amsmath,amsfonts,amssymb}
\usepackage{graphicx}
\usepackage{array}
\usepackage{wrapfig}
\usepackage{gensymb}
\usepackage[colorlinks=true, allcolors=blue]{hyperref}
\usepackage{aas_macros}
\usepackage{comment} 

\usepackage{float}
\graphicspath{ {./figures/} }

\title{Experimental demonstration of spectral linear dark field control at NASA's high contrast imaging testbeds}

\author[a]{Phillip K. Poon}
\author[a]{Axel Potier}
\author[a]{Garreth Ruane}
\author[a]{Alex B. Walter}
\author[a]{A~J~Eldorado~Riggs}
\author[a]{Matthew Noyes}
\author[a]{Camilo Mejia Prada}
\author[b]{Kyohoon Ahn}
\author[b, c, d]{Olivier Guyon}

\affil[a]{Jet Propulsion Laboratory, California Institute of Technology, 4800 Oak Grove Dr, Pasadena, CA 91109}
\affil[b]{Subaru Telescope, National Astronomical Observatory of Japan, National Institutes of Natural Sciences (NINS), 650 North A'ohoku Place, Hilo HI 96720, United States}
\affil[c]{College of Optical Sciences, University of Arizona, Tucson, AZ 87521, United States}
\affil[d]{Steward Observatory, University of Arizona, Tucson, AZ 87521, United States}

\authorinfo{Further author information: Send correspondence to Phillip K. Poon\\Phillip K Poon: E-mail:phillip.poon@jpl.nasa.gov {\\  Government sponsorship acknowledged. \copyright 2023. All rights reserved. }}

\begin{document} 
\maketitle

\begin{abstract}

In order to directly image Earth-like exoplanets (exoEarths) orbiting Sun-like stars, the Habitable Worlds Observatory coronagraph instrument(s) will be required to suppress starlight to raw contrasts to $\sim10^{-10}$. Coronagraphs use active methods of wavefront sensing and control (WFSC) such as pairwise probing (PWP) and electric field conjugation (EFC) to create regions of high contrast in the science camera image, called dark holes (DH). 

Due to the low flux of these exoEarths, long exposure times are required to spectrally characterize them. During these long exposures, the contrast in the DH will degrade as the the optical system drifts from its initial DH state. To prevent such contrast drift, a WFSC algorithm running in parallel to the science acquisition can stabilize the contrast. However, PWP cannot be reused to efficiently stabilize the contrast since it relies on strong temporal modulation of the intensity in the image plane, which would interrupt the science acquisition. The use of small amplitude probes has been demonstrated but requires multiple measurements from each science sub-band to converge. Conversely, spectral linear dark field control (LDFC) takes advantage of the linear relationship between the change in intensity of the post-coronagraph out-of-band image and small changes in wavefront in the science band to preserve the DH region during science exposures. 

In this paper, we show experimental results that demonstrate spectral LDFC stabilizes the contrast to levels of a few $10^{-9}$ on a Lyot coronagraph testbed which is housed in a vacuum chamber. Promising results show that spectral LDFC is able to correct for disturbances that degrade the contrast by more than 100$\times$. To our knowledge, this is the first experimental demonstration of spectral LDFC and the first demonstration of spatial or spectral LDFC on a vacuum coronagraph testbed and at contrast levels less than $10^{-8}$.

\end{abstract}

\keywords{Coronagraph, Dark Hole Maintenance, Wavefront Sensing and Control, Linear Dark Field Control}

\section{INTRODUCTION}\label{sec:intro} 

Direct imaging of Earth-like exoplanets orbiting Sun-like stars is challenging, mainly because the reflected light from these planets is approximately 10 billion times fainter than their host stars \cite{Traub_2010}. A coronagraph is an optical instrument designed to perform direct imaging of exoplanets by attenuating the starlight. The coronagraph must employ a wavefront sensing and control (WFSC) loop to correct for wavefront aberrations in order to establish regions of high contrast in the science image called dark holes (DH). In this scenario, the WFSC algorithms must be able to sense and correct for wavefront errors at the picometer level \cite{NASA_ExEp_Progress}. 

A typical example of WFSC for coronagraphs is the combination of pairwise probing (PWP) \cite{Giveon_2011} and electric field conjugation (EFC) \cite{Giveon_2007}. Pairwise probing places known shapes on the deformable mirror (DM) which modulates the science camera image. By applying pairs of positive and negative probes, the difference image can be used to estimate the unknown electric field \cite{Giveon_2007, Potier_2020}. The estimate from PWP is the input of the EFC algorithm which then outputs a correction command to the deformable mirrors \cite{Given_Formalism_2009}. This process, called ``digging the dark hole'', is executed iteratively until the desired contrast is reached. 

In practice, the telescope may need to point at a bright reference star to obtain the suitable SNR required to establish the DH contrast in a reasonable amount of time \cite{CGI_Public_2022}. Once the desired contrast level is achieved, the DM command is stored, and the telescope can then point at the fainter science target star. Due to the low flux of these exoplanets, the science exposure may take up to several days or longer. This requires the entire optical system to be extremely stable. In reality, slow changes to the wavefront, called quasi-static aberrations, will corrupt the contrast in the DH, requiring time and resources to point to another suitably bright star to re-establish the desired contrast. If the coronagraph can maintain the DH contrast without the need to point to a bright star, this capability will significantly improve science yield, reduce required observing time, and lower costs\cite{Pogorelyuk_2020}. Even in situations where the target star is bright enough, the use of traditional PWP interrupts the science acquisitions. Another advantage of DH maintenance techniques is the potential to relax wavefront stability requirements. 

A previously demonstrated set of methods employs an Extended Kalman Filter (EKF) that utilizes DM dithering for wavefront sensing (WFS) \cite{Pogorelyuk_2019, Redmond_2020}. This is followed by an EFC-based correction algorithm to sustain the DH. These methods depend on images captured at the science wavelength and need a number of these images to accurately determine the appropriate DM command for correcting quasi-static wavefront errors. Also, DM dithering compromises the DH's contrast, thereby interrupting the science exposure. The speed of the estimation algorithm is limited by the exposure time. Because these methods use the science wavelength, they require longer integration times to maintain deeper contrasts. Minimizing the dither amplitude is possible but comes at the expense of estimator performance.

Developing a technique to maintain the DH contrast without interrupting the science acquisition is therefore important. Linear dark field control (LDFC) is a DH maintenance technique that does not require any modulation of the DM during the WFSC loop. There are two varieties of LDFC: spatial and spectral. In both varieties of LDFC, since the bright field has significantly more starlight than the DH, the changes in intensity are linear with small changes in wavefront error. The advantage of the linear relationship allows the use of linear control algorithms for WFSC to maintain the DH contrast. As discussed in prior work, LDFC is only intended as a DH maintenance procedure: it is not used to establish the initial DH. 

Spatial LDFC, which was the first to be proposed \cite{Miller_2017}, senses changes in intensity in the spatial bright field to detect quasi-static wavefront errors; this has been experimentally demonstrated on the Subaru Extreme Adaptive Optics testbed \cite{Ahn_2022, Ahn_2022_a&a}. However, spatial LDFC can only be applied to stabilize the DH in a 180\degree~  field of view. Spectral LDFC is a related concept that uses the spectral bright field, which is the out-of-band light, called the sensing band, to perform wavefront sensing in the whole (360\degree) field of view \cite{Guyon_2017_Spectral_LDFC}. Either a separate camera, a dual-band camera, or an integral field spectrometer (IFS) can be used to measure the intensity changes in the sensing band. 

In this manuscript, we experimentally demonstrate spectral LDFC at contrast levels of a few $10^{-9}$. This demonstration shows that we can successfully correct disturbances that degrade the contrast by a factor of over 100$\times$. To our knowledge, this is the first experimental demonstration of spectral LDFC. We will also discuss a modal control framework applicable to LDFC, where control modes are derived from a singular value decomposition (SVD) of the measured linear response, and a leaky integrator tuned for each control mode allows for fine tuning of LDFC performance and stability.

\section{Methods}

\subsection{Terminology}

In this manuscript, we will use Normalized Intensity (NI) as a proxy for raw contrast. The NI is defined as the 2-D intensity of the on-axis Point Spread Function (PSF) with the Focal Plane Mask (FPM), Lyot Stop (LS), and Field Stop (FS) in the coronagraph, divided by the maximum intensity of the PSF for a source in the center of the DH. We denote it formally as $\text{NI}(\xi, \eta)$, where $\xi$ and $\eta$ are the cartesian coordinates in angular distance often given in units of $\lambda/D$. Both terms, ``contrast'' and ``NI'', may be used in this manuscript, but only NI is actually computed. Note that for small changes in the PSF, the NI is proportional to intensity. Therefore, when we discuss intensity, we are actually referring to the computation of the NI.

The science band is the wavelength range in which the NI is scored or tracked. Ultimately, we want to maintain the NI in the science band's scoring region and thus keep it from drifting. The sensing band refers to the wavelength range in which we sense changes to the NI (inside some correction region) and attempt to correct them.


\subsection{Testbed Hardware}

For these experiments, we used the Decadal Survey Testbed 2 (DST2) \cite{Noyes_2023}, located at NASA's High Contrast Tested Facility (HCIT). The DST2 is housed inside a large vacuum chamber to simulate the pressure and thermal environment of a space-based coronagraph. The chamber can reach a pressure of less than 0.5 mTorr and exhibits thermal stability of $\sim$10~mK. The optical system of DST2 is mounted on a carbon composite optical table, and the optical bench is mechanically isolated from the environment with three passive vibration isolators (MinusK CM-1). Both passive and active thermal control are employed to reduce thermal expansion that would otherwise lead to quasi-static aberrations. The vacuum chamber itself is housed in a large cleanroom.

In order to switch between the sensing band and the science band, a supercontinuum broadband laser (NKT Photonics SuperK FIANIUM \cite{Fianium_2023}) is fed through a photonic crystal fiber (PCF) supporting a large spectral bandwidth, to a variable bandwidth monochromator with a 440 nm wide tuning range and variable bandwidth \cite{Varia_2023}. The variable filter allows us to switch between spectral bands in approximately 1 second. We used a step-index single-mode fiber to connect the output of the monochromator to the input of the coronagraph, which leads to a large reduction in flux beyond 650 nm. Ideally, the separation between the science and sensing band should be as large as possible, but we were limited by the bandwidth of the single-mode fiber. For these experiments, we used a science band center wavelength of $\lambda_{sci} =~$530 nm and a sensing band center wavelength of $\lambda_{sen} =~$650 nm. Both bands have a fractional bandwidth of $\Delta \lambda/ \lambda$~=~0.01.

For these experiments, the DST2 was configured with a Lyot FPM. We also used a single $50 \times 50$ actuator microelectromechanical (MEMS) DM manufactured by Boston Micromachines. The physical field stop (FS) used for this experiment has an outer edge of approximately 12.4 $\lambda_{sci}/D$ and an inner edge of 2.5 $\lambda_{sci}/D$.

The science camera, which is used for acquiring the science and sensing band images, is an Andor NEO sCMOS cameras. This camera has a 6.5$\mu$m pitch $2160 \times 2560$ pixel array. This camera is not vacuum rated, which requires us to enclose the camera in a pressurized vessel. A broadband window is used to transmit light to the camera. 

\subsection{Creating the dark hole region}

\begin{figure}[htbp!]
\centering
\includegraphics[scale=0.65]{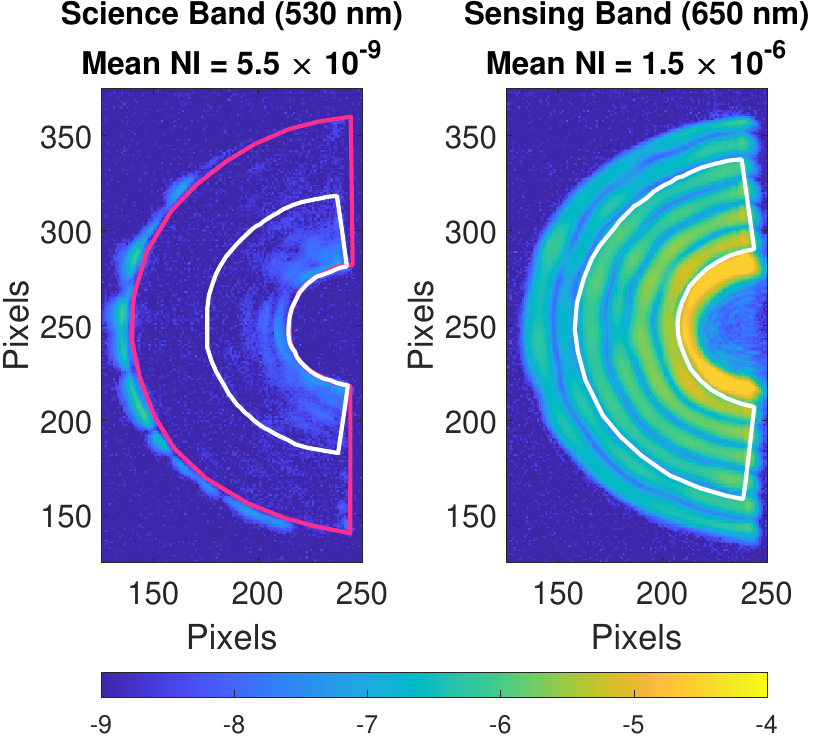}
\caption{
    The science band NI image (left) after creating a DH with several iterations of PWP and EFC. The red outline denotes the correction region used for EFC. The white outline denotes the science band mask. The sensing band NI image (right) for the same DM pattern. The white outline denotes the sensing band mask. Each image is the mean of ten exposures, each lasting 10 seconds. 
    \label{fig:unprobed_sci_and_sens_images}
}
\end{figure}

To create the DH, we used PWP and EFC which is implemented using a software library called FALCO \cite{Riggs_Falco_2018}, allowing us to reach NI levels of a few $10^{-9}$. The science band DH is shown in Fig.~\ref{fig:unprobed_sci_and_sens_images} (left), with the correction region (the designated region of pixels that will be brought to high contrast) shown inside the red outline. Once the desired NI in the science band DH is established, we store the DM command corresponding to this high contrast DH.

\subsection{Linearity in the sensing band}
\label{ss:linearity}

Both spatial and spectral LDFC are based on the assumption that in the small wavefront error regime, the changes in intensity in the bright field are linear with wavefront error\cite{Miller_2017, Ahn_2022_a&a, Guyon_2017_Spectral_LDFC}. For spectral LDFC, the bright field refers to the sensing band, where the mean NI is several orders of magnitude larger than in the science band (see in Fig.~\ref{fig:unprobed_sci_and_sens_images}, right ). 

To satisfy that we are in the linear range, we place sine wave patterns on the DM that correspond to localized speckle inside both the sensing and science band masks. We then vary the amplitude of the sine wave from -20 mV to 20 mV peak-to-valley (or approximately $\pm$0.2 nm PTV). The corresponding NI of the speckle generated on the focal plane is recorded and the NI from the unprobed image is subtracted.

\begin{equation}
    \Delta \text{NI} (\xi, \eta) = \text{NI}_p (\xi, \eta) - \text{NI}_{up} (\xi, \eta)
\end{equation}

\begin{figure}[htbp!]
\centering
\includegraphics[scale=0.90]{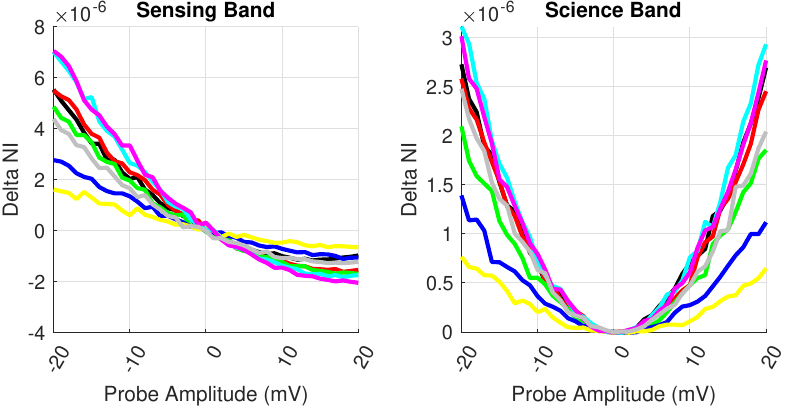}
\caption{
    Experimentally measured $\Delta$NI (NI of the probed image location - NI of unprobed image location) for several spatial locations in the sensing band (left) and the science band DH (right). The same color in both plots correspond to the same respective location in units of $\lambda/D$. This demonstrates the approximately linear response in the sensing band compared to the science band which exhibits a quadratic response. The $\pm$20~mV probe amplitude range corresponds to $\pm$0.2~nm~PTV in DM surface height. 
    \label{fig:linearity_test}
}
\end{figure}

Where $\text{NI}_p (\xi, \eta) $ is the normalized intensity image of the probed image and $\text{NI}_{up} (\xi, \eta)$ is the normalized intensity of the unprobed image. Figure \ref{fig:linearity_test}(right) clearly shows a quadratic response in the change of NI, $\Delta$NI, with respect to the sine wave pattern amplitude in the science band. In the sensing band, the response is approximately linear and monotonic over range of approximately -10 to 10 mV DM command (or approximately $\pm$0.1 nm PTV), as seen in Fig.~\ref{fig:linearity_test}(left).

\subsection{Calibration}
\label{ss:cal}

The spectral LDFC calibration procedure empirically measures the change in NI in the sensing band for each mode of the DM poke basis. The corresponding measurements are stored as columns in the $N \times M$ response matrix $\mathbf{R}$, where $N$ is the number of evaluation pixels in the sensing focal plane area and $M$ is the number of DM poke modes (not necessarily orthogonal at this point). 

Prior to beginning the calibration we generate the science and sensing band masks. First we specify the science band mask such that it is inside the DH region created with PWP and EFC. For the sake of this experiment, we choose a region whose edges range from 3.7 $\lambda/D$ to 8.0 $\lambda/D$ with an angular width of $165^{\circ}$. During real operation, that region should match the initial DH shape. 

\begin{figure}[htbp!]
\centering
\includegraphics[scale=0.60]{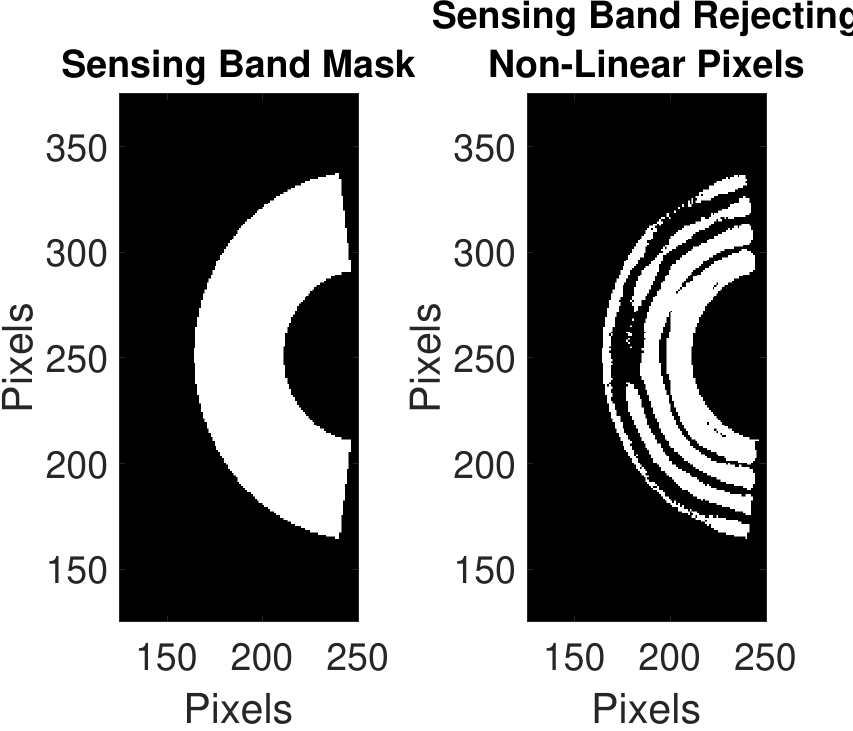}
\caption{
    Initial sensing band mask (left), the sensing band mask has an equivalent shape. The inner radius is 3.7 $\lambda/D$ and outer radius is 8 $\lambda/D$ with an angular width of $165^{\circ}$. The actual sensing band mask (right) is a subsample of the original sensing band pixel mask, designed to reject spatial locations with a non-linear pixel response. 
    \label{fig:pixel_masks}
}
\end{figure}

The initial sensing band mask shown in Fig.~\ref{fig:pixel_masks} (left) has the same dimensions as the science band as measured by angular coordinates $(\lambda/D)$. Therefore any speckles that we place on the sensing band will show up at the same location in the science band, scaled by their respective wavelength. 

In the second step, a non-linear pixel rejection mask is created (see in Fig.~\ref{fig:pixel_masks}, right). Here, we consider only the pixels inside the initial sensing band mask. Pixels whose NI values are lower than a  threshold times the median of the sensing band NI are rejected. Indeed, pixels with low NI in the unprobed image are more likely to exhibit a non-linear response to changes in wavefront. For the results shown in this paper we used a threshold value of 0.6.

The next step is to choose a DM poke basis for the calibration.  There are several possible bases to choose from such as the actuator basis (individual actuators), Hadamard basis, or sines wave patterns. We decided against using the actuator basis as it is time consuming and produces low SNR. Using the Hadamard basis increases the SNR but requires as many exposures as the actuator basis, which for our case is 2500. For these reasons we chose to use the sine wave basis which allows us to place high SNR speckle in the science and sensing band. We precomputed the sine wave DM patterns to overfill the sensing band pixel mask with speckle spaced 1 $\lambda/D$ apart in a square grid pattern.

During each iteration of the calibration loop, a positive and negative amplitude sine wave (positive and negative probe) is applied to the DM. Only measurements from the sensing band are required. The images from the positive probe and the negative probe are subtracted and scaled by the reciprocal of twice the amplitude:

\begin{equation}
    \mathbf{R}_m = \frac{1}{2A} \left( \text{ } \mathbf{n}_{m}^{+} - \mathbf{n}_{m}^{-}\text{ } \right), 
\end{equation}

\noindent where $\mathbf{n}_{m}^{+}$ is the vectorized sensing band image associated with the positive probe for the $m^{th}$ basis vector and $\mathbf{n}^{-}$ is associated with the negative probe, both are $N \times 1$ vectors. $A$ is the amplitude of the probe which in our case is 10 mV (approx. 0.1 nm PTV), to remain in the linear range (see in Sec.~\ref{ss:linearity}). The resulting vector $\mathbf{R}_m$ is placed in the $m^{th}$ column of $\mathbf{R}$. Therefore $\mathbf{R}$ is an $N \times M$ matrix, where $N$ is the number of pixels in the sensing band pixel mask and $M$ is the number of modes (basis vectors) used to during the calibration.

During each iteration of the calibration, we iteratively add to the reference measurement vector $\mathbf{n}_{ref}$ 

\begin{equation}
    \mathbf{n}_{ref} = \frac{1}{2M} \sum_{m = 1}^{M} \left( \mathbf{n}_{m}^{+} + \mathbf{n}_{m}^{-}  \right),
\end{equation}

\noindent where $\mathbf{n}_{ref}$ is an $N \times 1$ vector. 

\subsection{The wavefront sensing and control (WFSC) loop using modal control}
\label{sec:the_wfsc_loop}

In spectral LDFC, the forward model can be summarized as 

\begin{equation}
    \Delta \mathbf{n} = \mathbf{R} \Delta \mathbf{w},
\end{equation}

\noindent where $\Delta \mathbf{n}$ is the change in NI in the sensing band pixel mask and $\Delta \mathbf{w}$ is the change in wavefront relative to the initial DH wavefront. Thus, we can extend this concept to a WFSC loop by denoting the measurement in the $k^{th}$ iteration as

\begin{equation}
    \Delta\mathbf{n}_{k} = \mathbf{n}_{k} - \mathbf{n}_{ref},
    \label{eq:wfs_measurement}
\end{equation}

\noindent where each $\mathbf{n}_k$, an $N \times 1$ vector, is the vectorized measurement of NI inside the sensing band pixel mask at iteration $k$. 

One way to compute the correction command is to apply the pseudoinverse $\mathbf{R}^{\dagger}$. This can be computed using the singular value decomposition (SVD) as $ \mathbf{R}^{\dagger} = \mathbf{V} \mathbf{\Sigma}^{\dagger} \mathbf{U}^T$. Where $\mathbf{U}$ is an $N\times M$ matrix, $\mathbf{V}$ is an $M\times M$ matrix, and $\mathbf{\Sigma}$ is an $M\times M$ matrix. $ \mathbf{R^{\dagger}} \mathbf{ \Delta n}_k$ produces an $ M \times 1 $ correction vector in the original calibration basis. 

In practice, a threshold or weighting is placed on the singular values to prevent correcting noise, reduce cross-talk, or prune redundant dimensions \cite{Poon_2009}. Thresholding singular values has been successfully used for spatial LDFC \cite{Ahn_2022, Ahn_2022_a&a}. We use an extension of this technique which provides the ability to add gains and leaks to each eigenmode individually which we call the modal control approach \footnote{We do not claim any novelty to this approach.}. 

The modal control approach allows us to extend the SVD pseudoinverse for more precise control of each eigenmode during the spectral LDFC WFSC loop:

\begin{equation}
    \mathbf{x}_{k} = \mathbf{L} ( \mathbf{x}_{k-1} - \mathbf{G} \mathbf{\Sigma}^{\dagger} \mathbf{U}^{T} \Delta \mathbf{n}_{k}  ),
    \label{eq:X_k}
\end{equation}

\noindent where $\mathbf{x}_k$, an $M \times 1$ vector, is the control command per eigenmode. $\mathbf{G}$ is the modal gain, a diagonal matrix with elements $g_m$ and $\mathbf{L}$ is the modal leak, a diagonal matrix with elements $1 - l_m$. The leak values $l_m$ are a multiplicative factor used to drive commands to zero, implementing a leaky integrator control law.

To complete the loop and apply the correction to the DM, we compute the correction in the calibration basis we originally used to acquire the response matrix 

\begin{equation}
    \mathbf{f}_k = \mathbf{V} \mathbf{x}_k
    \label{eq:convert_to_sine_wave}
\end{equation}

\noindent where $\mathbf{f}_k$ an $M \times 1$ vector. Then multiply by $\mathbf{D}$ to convert to the actuator basis. Thus the control command in the actuator basis is

\begin{equation}
    \mathbf{p}_k = \mathbf{D} \mathbf{f}_k 
    \label{eq:convert_to_sine_wave_2}
\end{equation}

\noindent where the columns of $\mathbf{D}$ are the sine wave patterns from calibration. $\mathbf{D}$ is an $N_{act}^2 \times M$ matrix. $N_{act}$ is the number of actuators across the DM assuming a square actuator pattern. $\mathbf{p}_k$ is an $N_{act}^2 \times 1$ vector.

\begin{figure}
\centering
\includegraphics[scale=0.500]{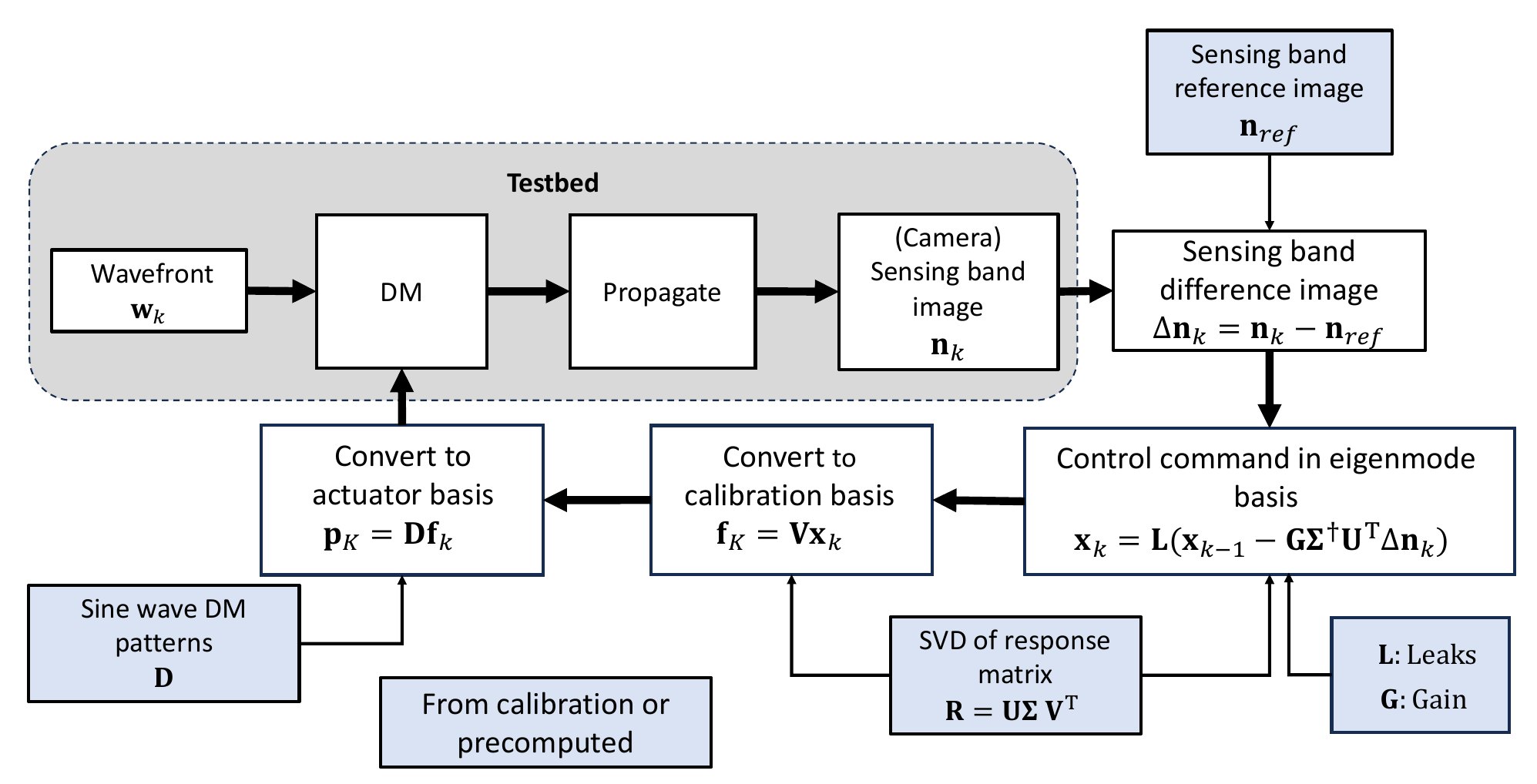}
\caption{
    The spectral LDFC wavefront sensing and control loop. Note there is no requirement to use the science band. The boxes with the gray area represent physical components or processes on the testbed. The light blue boxes are quantities that are precomputed or known prior to the spectral LDFC WFSC loop.  
    \label{fig:wfsc_loop}
}
\end{figure}

The spectral LDFC WFSC loop is summarized in Fig. \ref{fig:wfsc_loop}. This figure and equations \ref{eq:X_k}, \ref{eq:convert_to_sine_wave}, and \ref{eq:convert_to_sine_wave_2} show that the operations involved are mostly matrix multiplications.

\section{Experimental Results}

\begin{figure}[htbp!]
\centering
\includegraphics[scale=0.65]{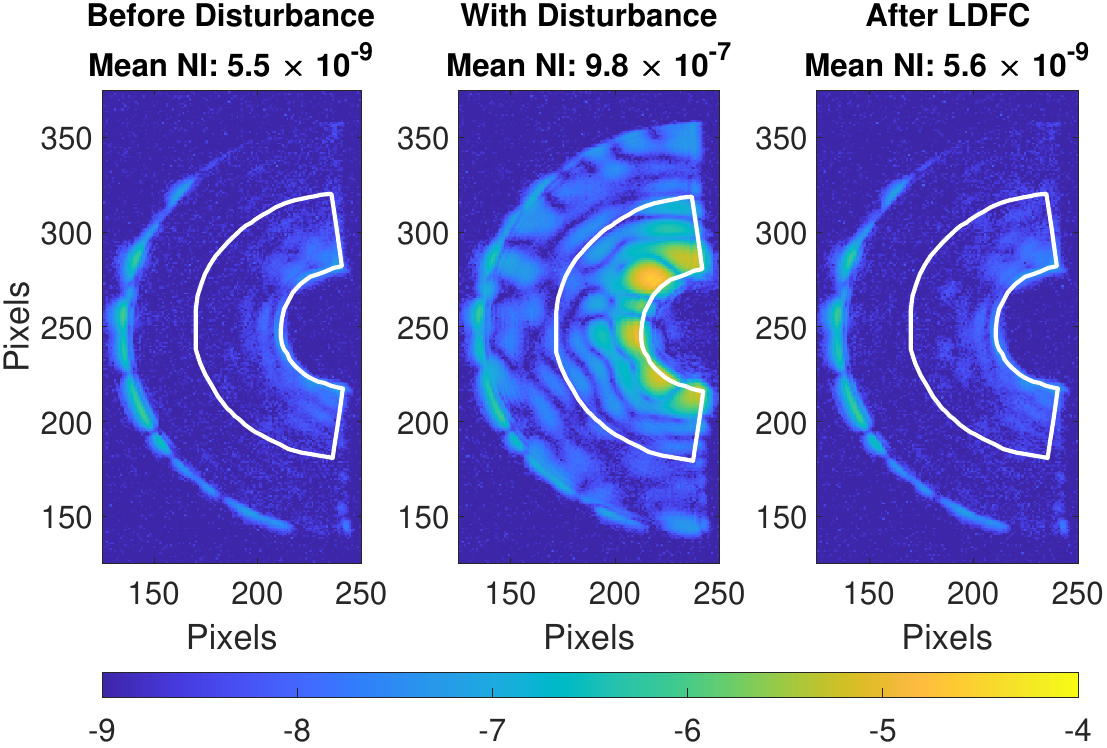}
\caption{
    The science band image ($\log_{10}(\text{NI})$) (left)~before a disturbance is injected using the DM and (center)~with a disturbance composed of the first seven eigenmodes applied to the DM . The mean NI over the science band mask is 100 times larger the initial mask. (Right) The science band image after 120 iterations of spectral LDFC and modal control, the science band image mean NI has been corrected to slightly larger than the original value of $5.5\times10^{-9}$. The white outline shows the science band pixel mask in which we use to score the mean NI in the science band. Each image is the average of 10 images at 10 second exposure time. The mean NI computed from the averaged image over the science band mask. 
    \label{fig:sci_band_imgs_before_and_after_ldfc}
}
\end{figure}

In order to demonstrate spectral LDFC on the DST2 testbed, we used PWP and EFC to establish a science band mean NI of $~5.5 \times 10^{-9}$ in the pixel mask (see in Fig.~\ref{fig:sci_band_imgs_before_and_after_ldfc}, left). We then applied a disturbance on the DM that corresponds to a linear combination of the first seven eigenmodes of the response matrix $\mathbf{v_1}, ..., \mathbf{v_7}$ each with an amplitude, $a_i$, of 20 mV (approx. 0.2 nm PTV). This is twice the amplitude that we used for the sine wave basis during the calibration. This degraded the mean NI in the science band pixel mask to $9.8 \times 10^{-7}$, as seen in Fig.~\ref{fig:sci_band_imgs_before_and_after_ldfc} (center).

\begin{figure}[htbp!]
\centering
\includegraphics[scale=0.70]{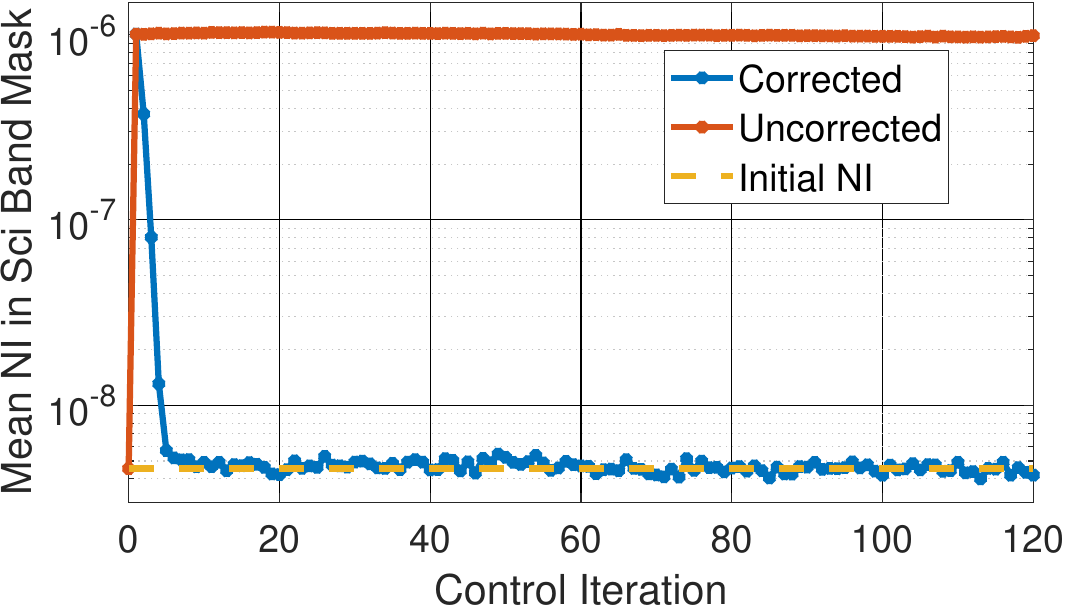}
\caption{
    Mean NI during the 120 iterations of spectral LDFC with modal control. The first seven eigenmodes of the response matrix are equally injected with the same amplitude, 20 mV, as a disturbance. The blue line is the mean NI in the science band mask with the correction as computed from modal control, applied to the DM. After 10 iterations the science band mean NI has been corrected. Spectral LDFC successfully detects and corrects the disturbances to nearly the original mean NI in the science band. The red line is the mean NI in the science band mask without any correction applied. The dashed yellow line shows the mean NI before the disturbance is injected to serve as a useful visual indicator.  
    \label{fig:sci_band_mean_ni_vs_itr}
}
\end{figure}

We then applied the spectral LDFC algorithm as explained in Sec.~\ref{sec:the_wfsc_loop}. Figure~\ref{fig:sci_band_mean_ni_vs_itr} shows a plot of both the closed-loop and open-loop performance. The red line is the mean NI in the science band mask without the correction applied. The blue line is the mean NI with the correction applied. The values shown in Fig.~\ref{fig:sci_band_mean_ni_vs_itr} are from a single exposure at 1 second integration times  for both the science and sensing band to reduce the duration of the WFSC experiment. After nine control iterations, the closed-loop mean NI improved to less than $10^{-8}$. After 120 control iterations, the corrected science band mean NI is slightly worse than the starting value, as seen in Fig.~\ref{fig:sci_band_imgs_before_and_after_ldfc} (right).  In the final iteration, the closed loop mean NI shows a factor of over 170$\times$ improvement compared the open loop mean NI.

\begin{figure}
\centering
\includegraphics[scale=0.65]{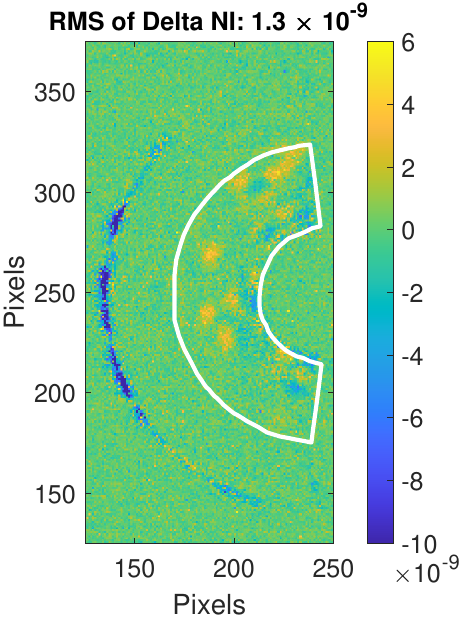}
\caption{
    The difference of the two science band averaged images from Fig.~\ref{fig:sci_band_imgs_before_and_after_ldfc}: after LDFC minus before disturbance, denoted as Delta NI (do not confuse this with $\Delta \mathbf{n}_k$ from Eq. \ref{eq:wfs_measurement}). The color scale is $\log_{10}$\text{Delta NI}. The root-mean-square (RMS) of the Delta NI in the science band mask is $1.3 \times 10^{-9}$. The RMS is computed over the science band mask, denoted by the white outline.
    \label{fig:sci_band_delta_ni}
}
\end{figure}

The difference between the after LDFC image and the before disturbance image is shown in Fig.~\ref{fig:sci_band_delta_ni}. To compute the inherent RMS in the difference image inside the science band mask one must remove the effect due to noise. The RMS is calculated as $\text{RMS}_\text{diff} = \left(\text{RMS}_\text{total}^2 - \text{RMS}_\text{noise}^2 \right)^{1/2}$, where $\text{RMS}_\text{diff}$ is the inherent RMS in the difference image, $\text{RMS}_\text{total}$ is the combined RMS from both the underlying difference image and noise, and $\text{RMS}_\text{noise}$ is the RMS of a $150 \times 150$ pixels corner where the FS is opaque. 

\begin{figure}
\centering
\includegraphics[scale=0.600]{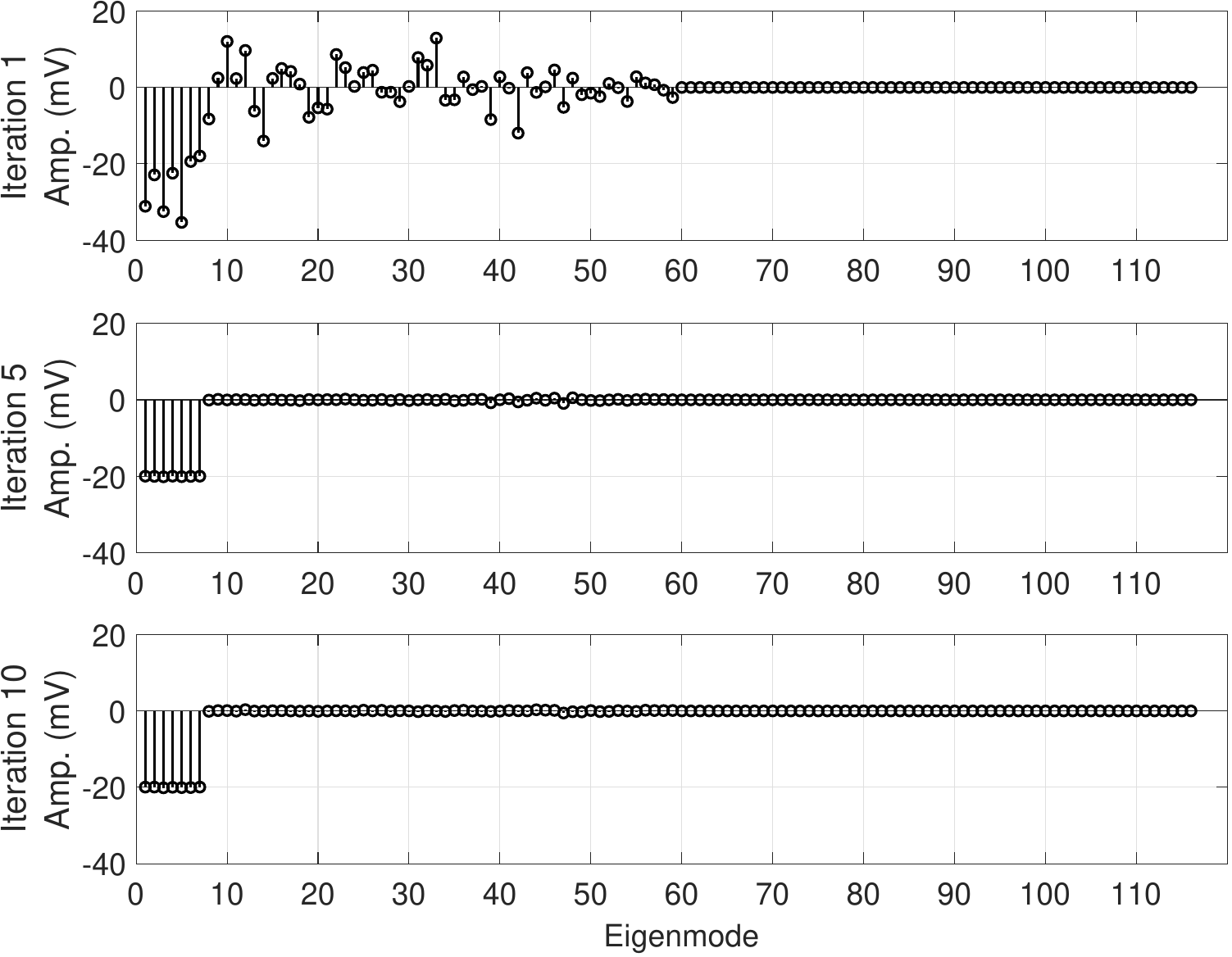}
\caption{
    (Top) The control for each eigenmode from equation \ref{eq:X_k} at iteration $k=1$. (Middle) The control at iteration $k=5$. (Bottom) The control at iteration $k=10$. The control is essentially opposite of the disturbance that we inject which is the first seven eigenmodes with amplitude 20 mV each. 
    \label{fig:correction_vs_eigenmode}
}
\end{figure}

We can monitor how the control (correction), produced by equation \ref{eq:X_k}, evolves with each WFSC iteration. The first iteration (see Fig.~\ref{fig:correction_vs_eigenmode}, top), over corrects the first eigenmode and other modes from 10 to 60. Modes 60 and higher have a leak $l_m$ of 1 and thus have no control applied. By the fifth iteration (see in Fig.~\ref{fig:correction_vs_eigenmode}, center), the first seven eigenmodes are being controlled to the correct value of -20 mV, with higher modes almost all being controlled to the correct value. After ten iterations (see in Fig.~\ref{fig:correction_vs_eigenmode}, bottom), the correction for each eigenmode is essentially the opposite of what we injected as a disturbance, canceling them out.

\subsection{Tuning the leaks}

\begin{figure}
\centering
\includegraphics[scale=0.90]{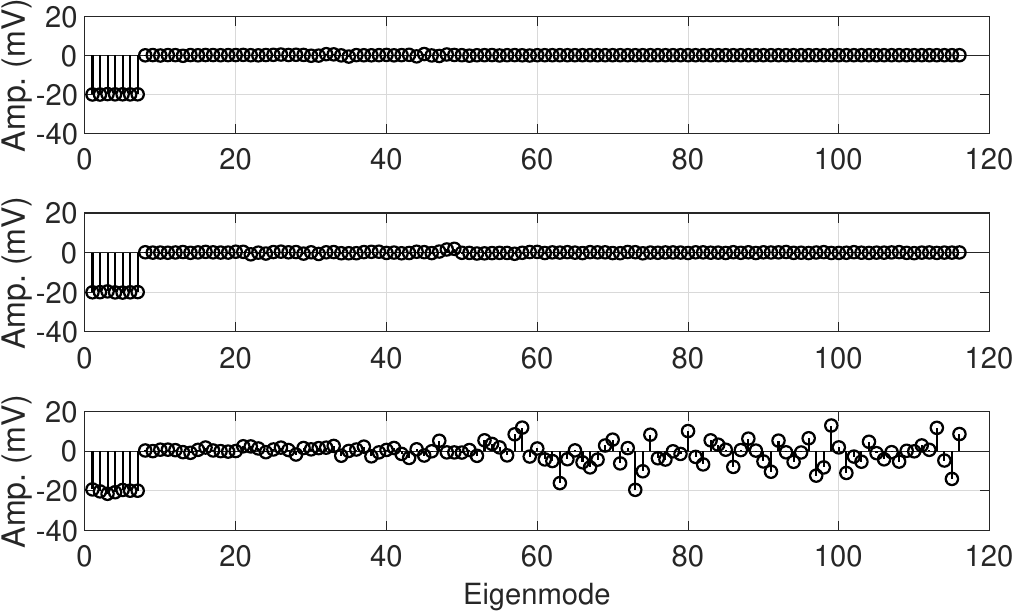}
\caption{
    (Top) The control for each eigenmode from equation \ref{eq:X_k} at iteration $k=10$ with a leak value of 0.6 for eigenmodes 50-59 and 1.0 for eigenmodes 60 and higher. (Middle) The control also at iteration $k=10$ but with a leak value set to 0.3 for eigenmodes 50-59 and 0.6 for eigenmodes 60 and higher. (Bottom) The control also at iteration $k=10$ but when all the leak values set to 0.
    \label{fig:control_vs_leak}
}
\end{figure}

\begin{figure}
\centering
\includegraphics[scale=0.70]{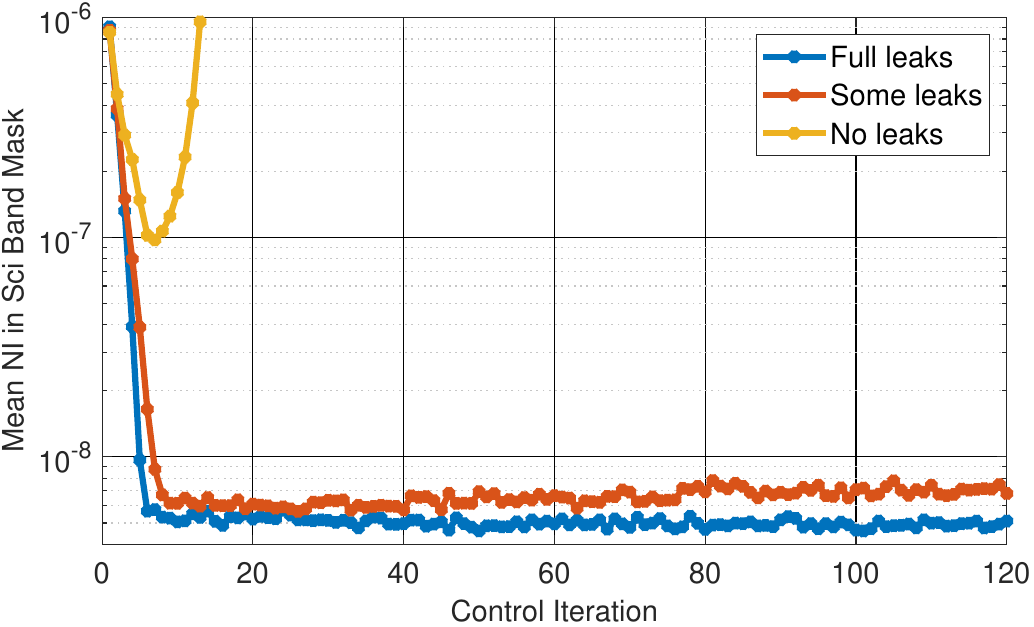}
\caption{
    (Blue) The mean NI in the science band for a leak value of 0.6 for eigenmodes 50-59 and 1.0 for eigenmodes 60 and higher. (Red) The mean NI in the science band for a leak value 0.3 for eigenmodes 50-59 and 0.6 for eigenmodes 60 and higher. (Yellow) The mean NI in the science band when all the leak values set to 0. 
    \label{fig:ni_vs_itr_for_various_leaks}
}
\end{figure}

We heuristically chose the leak values by observing how the lack of leaks affects the control (see Fig.~\ref{fig:control_vs_leak}, bottom). Without any leaks, the modal control algorithm attempted to control eigenmodes 50 and higher even though they were not added as synthetic disturbances. This caused the control loop to diverge after 6 iterations (see Fig.~\ref{fig:ni_vs_itr_for_various_leaks} yellow line plot). As we increased the strength of those leaks, both the control (see Fig.~\ref{fig:control_vs_leak}, middle and top) and the DH NI versus iteration became more stable (see Fig.~\ref{fig:ni_vs_itr_for_various_leaks} red and blue). 

For the results shown in Fig.~\ref{fig:sci_band_mean_ni_vs_itr} and \ref{fig:sci_band_imgs_before_and_after_ldfc}, we used leak values of $l_m = 0.6$ for eigenmodes 50 to 59 and turned off control ($l_m = 1.0$) for eigenmodes 60 to 116 (there are 116 eigenmodes). We noticed that eigenmodes 60 and higher were being controlled even though there was no disturbance being applied to them, this was causing the science band mean NI to drift. The higher eigenmodes tend to have lower SNR and higher cross-talk. Higher order eigenmodes also tend to have higher spatial frequency which are more difficult to realize on a deformable mirror.

\section{Conclusion}

We experimentally demonstrated spectral LDFC can sense and correct disturbances. Spectral LDFC depends on the fact that the intensity of out-of-band light varies linearly with small perturbations in wavefront error. We show this relationship is approximately true even though some pixels in the sensing band may have a non-linear response and must be rejected using a pixel mask.  More specifically, we demonstrated that we can inject a disturbance driving the NI to 170 times the original mean NI of $5.5 \times 10^{-9}$ and correct it. The change of intensity in the sensing band was used as input to a singular value decomposition based modal control algorithm to successfully correct for disturbances in the science band. A leaky integrator was also employed in the modal control algorithm to increase the stability of the spectral LDFC loop.

These are the first spectral  LDFC results to be demonstrated at NI levels of a few $10^{-9}$ on a vacuum testbed which is designed to simulate the environment of a space-based coronagraph. 

Future work will focus on practical aspects of using spectral LDFC for DH maintenance. We plan to show that spectral LDFC can be used to control for the natural drift of a coronagraph testbed. Another open question is understanding what kinds of disturbances cannot be corrected with spectral LDFC. For example, a calibration based on sine waves will mostly not be able to correct for speckle created by a cosine. Therefore we may need to add extra or different kinds of vectors to our basis to enlarge the set of disturbances we can correct for. Especially relevant to answering the question of the need to point a telescope to a bright reference star is the ability to execute spectral LDFC in the low photon flux regime. 

Alternative strategies to spectral LDFC for active DH maintenance have already been published \cite{Redmond_2020, Ruane_DPLC_2023}. We intend to compare the performance of these strategies on the same high contrast imaging testbed. 

\acknowledgments 

The research was carried out at the Jet Propulsion Laboratory, California Institute of Technology, under a contract with the National Aeronautics and Space Administration (80NM0018D0004).

\bibliography{report} 
\bibliographystyle{spiebib} 

\end{document}